\documentclass[twocolumn,showkeys,showpacs,prb]{revtex4}
\usepackage{amsfonts, amsmath, latexsym}
\usepackage[dvips]{graphicx}

\newtheorem{1}{Proposition}
\newtheorem{2}[1]{Proposition}
\newtheorem{3}{Theorem}
\newtheorem{4}[3]{Theorem}
\newtheorem{5}[3]{Theorem}

\begin{document}

\title{Reconstruction of thermally-symmetrized quantum autocorrelation functions from imaginary-time data}

\author{Cristian Predescu}
\email{cpredescu@comcast.net} 
\affiliation{Department of Chemistry and Kenneth S. Pitzer Center for Theoretical Chemistry, University of California, Berkeley, California 94720}

\date{\today}
\begin{abstract}
In this paper, I propose a technique for recovering quantum dynamical information from imaginary-time data via the resolution of a one-dimensional Hamburger moment problem. It is shown that the quantum autocorrelation functions are uniquely determined by and can be reconstructed from their sequence of derivatives at origin. A general class of reconstruction algorithms is then identified, according to Theorem~3. The technique is advocated as especially effective for a certain class of quantum problems in continuum space, for which only a few moments are necessary. For such problems, it is argued that the derivatives at origin can be evaluated by Monte Carlo simulations via  estimators of finite variances in the limit of an infinite number of path variables. Finally, a maximum entropy inversion algorithm for the Hamburger moment problem is utilized to compute the quantum rate of reaction for a one-dimensional symmetric Eckart barrier. 
\end{abstract}
\maketitle

\section{Introduction}
While providing a formally simple solution for the quantum dynamics of a physical system, the Feynman path integral method\cite{Fey48} generates one of the most difficult problems in computational physics, when it comes to the actual simulation on a ``classical'' computer: the dynamical sign problem.\cite{Ami93, Ber86} The highly oscillatory integrals appearing in the Feynman path integral expression of the propagator cannot be computed by direct Monte Carlo techniques, for there is no suitable importance function that would transform the propagator into an integral against a probability distribution.\cite{Cam60} Despite this inherent limitation, the Monte Carlo methods have been applied to the finite-temperature dynamics,  through two different approaches, mainly. Both techniques alleviate only partly the exponential loss of signal that is indicative of the dynamical sign problem. The first approach tries to construct an appropriate importance function by  convoluting the highly oscillatory integrand with a local distribution probability, whether a continuous\cite{Dol84, Fil86, Mak87, Dol88} or a discrete one.\cite{Mak92} While the research continues, actual applications of such techniques to realistic physical systems are, at present, rare.

The second approach, of which the present development is part, attempts to reconstruct certain dynamical correlation functions from the imaginary-time counterparts. While analytical continuation arguments show that this is possible in principle,\cite{Bay61, Nel64} the resulting algorithms always involve the resolution of certain ill-posed  numerical problems, as for instance inverse real Laplace transforms or inverse moment problems. Such inverse problems are highly unstable and suffer from an exponential amplification of the errors in the input data. They require very accurate Monte Carlo data, a careful choice of the type of data that is computed, as well as appropriate choices of inversion and regularization algorithms. 

The most common strategy for performing the analytic continuation is based on the inversion of a two-sided real Laplace transform  with noisy input data, to recover a spectral function.\cite{Sch85, Whi89, Jar89, Jar96} As already mentioned before, the inversion problem is ill-posed and, as a consequence, the inversion algorithms are highly unstable. The lack of continuity of the inverse Laplace transform with the input data causes any inversion algorithm to amplify the errors in the input data in an exponential fashion. Because these errors are of statistical nature, one is inclined to believe that it is virtually impossible to recover any useful dynamical information. However, most notably by use of methods of Bayesian statistical inference with entropic priors,\cite{Gub91, Jar96} various research groups have been successful in obtaining limited but useful and sometimes surprisingly reliable quantum dynamical information, whether in the form of spectra,\cite{Thi83, Gal94, Kim97, Kim98, Kri01} quantum rates of reaction,\cite{Mil83, Rab00, Sim01} or diffusion constants.\cite{Rab02}

The Monte Carlo data computed for the methods based on the inverse Laplace transform are usually the values on a grid of some imaginary-time correlation function,\cite{Jar96} values that have proportional statistical errors and are highly redundant. However, to a larger extent, the quality of the reconstructed spectral density is controlled by the errors of the relative differences, in addition to the errors of the absolute values of these data. It is then apparent that ensuring low relative errors for such differences is a definite way of improving the quality of the final results, as well as the stability of the inversion algorithms. Depending on the order $k$ of the finite-difference scheme considered, the value of such a difference decreases as a polynomial of order $k$ with the mesh of the grid, and so must decrease its error. It is then apparent that good quality of the input data requires good error bars, not only for the imaginary-time correlation functions, but also for their high-order derivatives. At extreme, one may consider that the input data consist of the value of the imaginary-time correlation function at time zero only,  together with the sequence of derivatives at origin. If such data are computed, the reconstruction of the spectral density involves the resolution of an inverse moment problem, as we shall discuss in the next section. The values of the derivatives at origin of the imaginary-time correlation function become the moments of the spectral density, except perhaps for a normalization coefficient.  The inverse moment problem is expected to be more stable than the inverse Laplace transform, with respect to the relative errors in the input data. 

It is quite unfortunate that, for general quantum problems, especially those in continuum space, the computation of derivatives at origin of imaginary-time correlation functions is extremely difficult. When available, the moment information can act as a stabilizing factor for the techniques based on the inverse Laplace transform. For instance, White\cite{Whi91, Whi92} utilized the first two moments of the spectral function for the two-dimensional Hubbard model as additional constraints in a maximum entropy approach, with remarkable success.  For Fermionic systems, Caffarel and Ceperley\cite{Caf92} utilized the first moment (average energy) of the spectral overlap function, moment evaluated by quantum Monte Carlo,  to stabilize their maximum entropy computations. Another way to make use of a limited number of low-order moments is to incorporate the information into the default model. This technique was utilized by Diesz and coworkers to reconstruct spectral weight functions for the one-dimensional t-J\cite{Dei92} and Heisenberg\cite{Dei93} models.

The limited use of moment information in the examples mentioned in the preceding paragraph is due to the difficulties encountered in the actual computation of the moments: for example, only low-order moments can be computed in an efficient way  by means of sum rules. In principle, when moments can be computed effectively, full-fledged moment techniques may be developed. However, as far as the present author is aware, this is only the case for the computation of moments for certain sparse Hamiltonian matrices, moments that can be computed in O(N) operations by stochastic methods, as shown by  Skilling,\cite{Ski89} as well as by Silver and R{\"o}der.\cite{Sil94} The availability of such information has led various groups to the application of Bayesian inference methods,\cite{Dra93} kernel polynomial methods,\cite{Sil94, Wan94, Sil96} or both\cite{Sil97} to the development of linear scaling algorithms for the resolution of densities of states, in electronic structure calculations.

There are a couple of problems in continuum space  that would particularly benefit from a moment approach. For these problems, the spectral function can be made rather featureless and the number of necessary moments can be made rather small by the variation of certain physical parameters. One interesting case is the aforementioned problem of computing the Fermion ground state by quantum Monte Carlo,\cite{Caf92} where the complexity of the spectral overlap can be greatly reduced by a proper choice of the  antisymmetric trial function (this spectral density becomes a single delta function, in the limit that the exact antisymmetric trial function is utilized). In chemical physics, a very important problem is the computation of the quantum rate of reaction by time-integrating the flux-flux correlation function associated with a surface that divides the reactants from the products.\cite{Mil74, Mil83, Mil98} It is known that the quantum rate of reaction does not depend upon the specific choice of dividing surface, although the complexity of the flux-flux correlation function is strongly dependent upon such a choice.\cite{Mil98} Thus, provided that ``the right'' choice of surface is made,  the spectral density of the flux autocorrelation function can be recovered effectively from a few moments, at least in principle. The only requirement is that these moments, or, equivalently, the derivatives at origin of the imaginary-time flux-flux correlation function, be computed with sufficient precision. A path-integral technique that shall be presented in Section~III is advocated as an effective way to compute derivatives at origin of correlation functions, for the type of problems discussed in the present paragraph. 

The first part of the paper provides a formal proof that the sequence of moments uniquely determines the spectral density and, therefore, the autocorrelation function. In addition, a convergence result is proved in order to identify a class of reconstruction algorithms for the autocorrelation functions. This result, which is the statement of Th.~\ref{Th:3}, demonstrates that all algorithms that preserve the positivity of the underlying spectral function (such algorithms are said to be positivity-preserving),\cite{Ath02} and which exactly match the first $n$ moments, lead to the correct autocorrelation function, in the limit $n \to \infty$. The algorithms based on the maximum entropy principle\cite{Jay57, Jar96, Tag94} as well as those based on kernel density functions\cite{Ath02, Sil94, Wan94} are examples of such algorithms.  Although the proofs are conducted for thermally-symmetrized autocorrelation functions, Th.~\ref{Th:3} applies for all correlation functions, the power spectra of which are positive distributions.

The larger part of the paper is concerned with the computation of  derivatives at origin of correlation functions for problems in continuum space. A general strategy for developing estimators having finite variance in the limit of an infinite number of path variables is discussed and illustrated for the case of the flux autocorrelation function. This strategy follows the general guidance of Predescu and Doll of burying the time dependence of paths into the potential part of the Feynman-Kac formula.\cite{Pre02} As implemented in the present paper, the computation of the derivatives requires the utilization of finite-difference schemes.  Such an approach has been successfully utilized in recent work for the numerical evaluation of several thermodynamic energy and heat-capacity estimators.\cite{Pre03} It is imperative to mention that \emph{no} differentiation of Monte Carlo data is ever attempted. Rather, the finite-difference scheme replaces the analytical evaluation of the derivatives of a deterministic function, evaluation that leads to expressions involving a large number of high-order partial derivatives of the potential, if performed.  The additional problem we must face in the present paper is that the utilization of  finite-difference schemes for derivatives beyond a certain order requires extended precision arithmetic, which may be a serious programming nuisance. Alternative techniques, as for instance Lyness' method,\cite{Lyn68} are possible, but they require analytic continuation of the potential in $d$-dimensional complex spaces. Such alternatives will be investigated in future work.

The present paper is limited to demonstrating that the advocated technique actually works.  Any issues of efficiency are postponed for future studies. In particular, these studies will have to address  the scaling of the variance of the estimators utilized for the computation of the moments with the order of the derivatives, the dimensionality of the system, and the inverse temperature. However, the numerical results presented in Section~IV (these results are quantum rates for a symmetric Eckart barrier)  show that the technique discussed in the present paper is a useful tool for obtaining quantum information of known computational difficulty. 

\section{The reconstruction of autocorrelation functions as an inverse moment problem}

In physics, the quantum dynamical information measured in experiments can generally be expressed in terms of quantum correlation functions of the type
\begin{equation}
\label{eq:1}
C_{O}(t) = \frac{\mathrm{tr}\left(e^{-\beta H}O^\dagger e^{iHt/\hbar}Oe^{-iHt/\hbar}\right)}{\mathrm{tr}\left(e^{-\beta H}\right)}, \quad t \in \mathbb{R},
\end{equation}
whenever the linear response theory provides a good approximation of the  measuring physical  process.\cite{Dol99r} The operator $H$ stands for the Hamiltonian of the system,  a self-adjoint and bounded from below operator, whereas $t \in \mathbb{R}$ and $\beta = 1/(k_B T) > 0$ are the real time and the inverse temperature, respectively. $O^\dagger$ denotes the adjoint of the operator $O$. 

The normalization term $\mathrm{tr}\left(e^{-\beta H}\right)$ in Eq.~(\ref{eq:1}) is not relevant for our discussion and we drop it from now on.  Using trace invariance in Eq.~(\ref{eq:1}), we obtain
\begin{equation}
\label{eq:2}
C_{O}(t) = \mathrm{tr}\left[e^{-(\beta  + i t / \hbar)H }O^\dagger e^{iHt/\hbar}O\right],  \quad t\in \mathbb{R}.
\end{equation}
Eq.~(\ref{eq:2}) is mathematically well-defined on the strip of the complex plane determined by the equation $0 < \mathrm{Im}(t) < \beta \hbar$, provided that
\begin{equation}
\label{eq:3}
\mathrm{tr}\left(e^{-\beta_1  H}O^\dagger e^{-\beta_2 H}O\right) < \infty,
\end{equation}
for all $\beta_1, \beta_2 > 0$. In these conditions, Baym and Mermin\cite{Bay61} have argued that the function $C_{O}(t)$ is analytic on the aforementioned domain. In addition,  $C_{O}(t)$ is uniquely determined on this domain by the values of $C_{O}(t)$ on the purely imaginary interval $(0, i\beta \hbar)$, values that can be computed efficiently by path-integral Monte Carlo techniques (via the Feynman-Kac formula). Finally,  the correlation function $C_{O}(t)$ is uniquely determined at all points of continuity on the frontier of the strip $0 \leq \mathrm{Im}(t) \leq \beta \hbar$, frontier that obviously includes the real axis. 

Berne and Harp\cite{Ber70} have pointed out that the computation of thermally-symmetrized quantum correlation functions 
\begin{equation}
\label{eq:4}
G_{O}(t)= \mathrm{tr}\left(e^{-\overline{\beta_c} H}O^\dagger e^{-\beta_c H }O\right),
\end{equation}
with $\beta_c = \beta /2 + it/\hbar$ and $\overline{\beta_c} = \beta /2 - it/\hbar$, might be an easier computational task, yet the correlation functions $G_{O}(t)$ and $C_{O}(t)$ carry essentially the same information, because their Fourier transforms satisfy the simple relation
\[
\bar{G}_{O}(\omega) = e^{-\beta \hbar \omega / 2}\bar{C}_{O}(\omega).
\]
Certain quantities of physical interest may not even require the computation of  direct and inverse Fourier transforms. For the flux (F) autocorrelation functions, Miller, Schwartz, and Tromp\cite{Mil83} have shown that the time integrals over the interval $[0,\infty]$ of the $C_{F}(t)$ and $G_{F}(t)$ functions are equal and so, for the determination of the quantum rate of reaction,\cite{Mil74, Mil83, Mil98} it does not matter which of the correlation functions is utilized.  As Eq.~(\ref{eq:3}) implies, $G_{O}(t)$ is well-defined in the  strip of the complex plane defined by the equation $|\mathrm{Im}(t)| < \beta \hbar / 2$. Baym and Mermin's argument can by extended to justify that $G_{O}(t)$ is analytic in this strip and admits a unique analytic continuation from the values of  $G_{O}(t)$ on the purely imaginary interval $(-i\beta \hbar / 2, i\beta \hbar / 2)$.

Before continuing with our exposition, let us remember the statement of the Hamburger moment problem. Suppose a sequence of real positive numbers $\{\mu_k \geq 0; k \geq 1\}$ is given. The Hamburger moment problem consists in answering the following questions:
\begin{enumerate}
\item{Is there a probability distribution $dP(\omega)$ on the real axis $(-\infty, \infty)$ such that
\[
\mu_k = \int_\mathbb{R} \omega^k dP(\omega), \quad \forall \; k \geq 1?
\]
}
\item{If the answer is positive, is the solution unique? (In this case, the problem is called determinate.)}
\item{If the solution is not unique, can one describe all possible solutions having moments $\mu_k$?}
\end{enumerate}
As a historical note, the moment problem on the interval $[0,\infty)$ is called a Stieltjes problem, whereas the moment problem on a compact interval $[a,b]$ is called the Hausdorff moment problem. The problems are called after the names of the mathematicians that have successfully and completely resolved the respective problems (the conditions are slightly different for the three cases, with the Hamburger problem being the most restrictive and challenging of the three). For our purposes it suffices to notice that a determinate Hamburger solution, if it is a Stieltjes or Hausdorff solution, then it is also determinate in the sense of Stieltjes or Hausdorff. Necessary and sufficient conditions for a sequence of positive numbers to be a moment sequence have been given by Hamburger in a series of papers from 1920 to 1921.\cite{Ham20} He has also produced sufficient and necessary conditions for the problem to be determinate. Because the inverse Hamburger moment problem lacks continuity with the input data,  any inversion algorithm designed to recover a probability distribution from moment data is ill-conditioned.

In this section,  the computation of thermally-symmetrized quantum correlation functions is reduced to an inverse Hamburger moment problem. In this respect, it is first shown that the sequence of derivatives at origin of the autocorrelation function is a sequence of moments $\{\mu_k; k \geq 0\}$, up to a normalization factor. In fact, the quantum autocorrelation function is the characteristic function of the probability distribution from which the moments $\mu_k$ are derived, probability distribution that is commonly called the spectral weight function. The ensuing Hamburger moment problem is then shown to be determinate. Therefore, the autocorrelation function is uniquely determined by its sequence of derivatives at origin. While this statement also follows from the Baym and Mermin's argument, our proof has the advantage of also suggesting reconstruction techniques. As such, Th.~\ref{Th:3}, which identifies a large class of candidate algorithms for the inverse Hamburger moment problem, does not follow from Baym and Mermin's argument.

\subsection{The input data}

Let us show that $G_O(t)$ is well-defined on the strip $|\mathrm{Im}(t)| < \beta \hbar / 2$ in the complex plane, whenever
\begin{equation}
\label{eq:7}
M_O(\beta_1,\beta_2) = \mathrm{tr}\left(e^{-\beta_1 H}O^\dagger e^{-\beta_2 H }O\right) < \infty,
\end{equation}
for all $\beta_1, \beta_2 > 0$. With the help of the spectral decomposition 
\begin{equation}
\label{eq:8}
e^{-\beta H} = \int_{\mathbb{R}} e^{-\beta E} |E\rangle \langle E | dE,
\end{equation}
Eq.~(\ref{eq:4}) becomes
\begin{equation}
\label{eq:9}
G_{O}(t)= \int_\mathbb{R} \int_\mathbb{R} e^{-\overline{\beta_c} E - \beta_c E'} |\langle E| O | E' \rangle|^2dEdE',
\end{equation}
whereas the condition given by Eqs.~(\ref{eq:3}) or (\ref{eq:7}) now reads
\begin{equation}
\label{eq:10}
M_O(\beta_1, \beta_2) = \int_\mathbb{R} \int_\mathbb{R} e^{-\beta_1 E - \beta_2 E'} |\langle E| O | E' \rangle|^2dEdE' < \infty, 
\end{equation}
for all $\beta_1, \beta_2 > 0$. From the inequality
\begin{eqnarray}
\label{eq:10a} \nonumber
|G_O(t)| \leq \int_\mathbb{R} \int_\mathbb{R} \left|e^{-\overline{\beta_c} E - \beta_c E'}\right| |\langle E| O | E' \rangle|^2dEdE' \\ = \int_\mathbb{R} \int_\mathbb{R} e^{-\beta(E + E')/2} |\langle E| O | E' \rangle|^2dEdE' = G_O(0), 
\end{eqnarray}
one concludes that the integral appearing in Eq.~(\ref{eq:9}) is absolutely convergent for all $t \geq 0$, because $G_{O}(0) = M_O(\beta/2, \beta/ 2) < \infty$. 
 
In these conditions, Baym and Mermin have argued that $G_O(t)$ is analytic on the strip $|\mathrm{Im}(t)| < \beta \hbar / 2$. In fact, the analyticity of the autocorrelation function follows easily from Eq.~(\ref{eq:7}) and from the absolute convergence of the integral appearing in Eq.~(\ref{eq:9}). Nevertheless, for our algorithm, we only need analyticity at origin together with a stronger statement on the radius of convergence of the Taylor series about origin. This is ensured by the following proposition.
\begin{1}
\label{Pr:1}
$G_O(t)$ is differentiable at origin infinitely many times and the radius of convergence of the Taylor series about origin is at least $\beta \hbar /2 $.  
\end{1}

\emph{Proof.} 
Consider the standard inequality 
\begin{eqnarray*}
\left|e^z - \sum_{k = 0}^{n-1} \frac{z^k}{k!} \right| \leq \sum_{k = n}^\infty \frac{1}{k!} |z|^k = \sum_{k = n}^\infty \frac{1}{k!} \frac{|z|^k}{r^k} r^k  \\ \leq \left({|z|}/{r}\right)^n \sum_{k = n}^\infty \frac{1}{k!} r^k \leq \left({|z|}/{r}\right)^n e^r,
\end{eqnarray*}
which is valid for all $|z| \leq r < \infty$. Pick an arbitrary positive number $r < \hbar \beta /2$. Then, for all $t$ with $|t| < r$, we have
\begin{eqnarray*}
\left|e^{it(E-E')/\hbar}- \sum_{k=0}^{n-1} \frac{1}{k!} \left(\frac{it}{\hbar}\right)^k(E-E')^k \right| \\ \leq \left({|t|}/{r}\right)^n e^{r |E-E'|/\hbar}.
\end{eqnarray*}
The last inequality implies
\begin{equation}
\label{eq:11}
\left| G_O(t) - \sum_{k=0}^{n-1} \frac{(it)^k}{k!} D_k  \right| \leq \left(\frac{|t|}{r}\right)^n M_r, 
\end{equation}
where 
\begin{equation}
\label{eq:12}
D_k = \frac{1}{\hbar^k}\int_\mathbb{R} \int_\mathbb{R} e^{-\beta(E+E')/2} \left(E-E'\right)^k |\langle E| O | E' \rangle|^2dEdE'
\end{equation}
and 
\begin{eqnarray*}
M_r = \int_\mathbb{R} \int_\mathbb{R} e^{-\beta(E+E')/2} e^{r|E-E'|/\hbar} |\langle E| O | E' \rangle|^2dEdE' \quad \ \\ 
= 2 \int_\mathbb{R} dE \int_{E}^\infty dE' e^{-(\beta/2 - r/\hbar)E' - (\beta / 2 + r / \hbar) E} |\langle E| O | E' \rangle|^2 \\  \leq 2 M_O(\beta/2 - r/\hbar, \beta/2 + r/\hbar) < \infty.
\end{eqnarray*}
The finitude of the last term follows from Eq.~(\ref{eq:7}) because $0 < r < \beta \hbar / 2$.

Since $M_r$ does not depend upon $n$, an easy inductive argument over $n$ and Eq.~(\ref{eq:11}) show that the terms $D_k$ are finite. Moreover, letting $n \to \infty$ in Eq.~(\ref{eq:11}), we learn that
\begin{equation}
\label{eq:13}
G_O(t) = \sum_{k = 0}^\infty \frac{1}{k!} (it)^k D_k
\end{equation}  
for all $t$ with $|t| < r$. Since $r < \beta \hbar / 2$ is arbitrary, the proof is concluded. \hspace{\stretch{1}}$\Box$

From Eq.~(\ref{eq:9}), we notice that $G_0(t) = G_0(-t)$. Therefore, $D_{2k+1} = 0$ for all $k \geq 0$. In these conditions, a little thought shows that Eq.~(\ref{eq:13}) can also be written as
\begin{equation}
\label{eq:14}
G_O(it) = \sum_{k = 0}^\infty \frac{1}{(2k)!} t^{2k} D_{2k},
\end{equation}  
the right-hand side series being convergent at least on the disc of equation $|t| < \hbar \beta / 2$. Thus, the numbers $D_{2k}$ are positive [by Eq.~(\ref{eq:12})] and are the even derivatives of the imaginary-time correlation function $G_O(it)$.

To summarize, the input data for the algorithm considered in the present paper is the sequence of even derivatives of the imaginary-time autocorrelation function $G_O(it)$. This sequence, denoted by $D_{2k}$, consists of positive numbers computable by path-integral Monte Carlo simulations.

\subsection{The function that is reconstructed}

The function (distribution) that is reconstructed is the power spectrum of the auto-correlation function $G_O(t)$. The power spectrum is defined through the identity
\begin{equation}
\label{eq:15}
\bar{G}_O(\omega)=\frac{1}{2\pi} \int_{\mathbb{R}} e^{-i\omega t} G_O(t) dt
\end{equation}
and is generally defined as a non-negative tempered distribution. 
With the help of Eq.~(\ref{eq:9}), one computes
\begin{eqnarray*}
\bar{G}_O(\omega) = \int_\mathbb{R} \int_\mathbb{R} e^{-\beta (E + E')/2} \left[\frac{1}{2\pi} \int_{\mathbb{R}} e^{it[-\omega + (E - E')/\hbar]} dt \right] \\ \times |\langle E| O | E' \rangle|^2dEdE' = \int_\mathbb{R} \int_\mathbb{R} e^{-\beta (E + E')/2} \\ \times \delta[-\omega + (E - E')/\hbar ] |\langle E| O | E' \rangle|^2dEdE'.
\end{eqnarray*}
Simple manipulations lead to
\begin{equation}
\label{eq:16} 
\bar{G}_O(\omega) = \hbar e^{- \beta \omega \hbar /2}\int_\mathbb{R}  e^{-\beta E } |\langle E+\omega \hbar| O | E \rangle|^2dE,
\end{equation}
which shows that the power spectrum is a non-negative distribution.

By means of Eq.~(\ref{eq:15}), one easily proves that the symmetry of $G_O(t)$ implies the symmetry of $\bar{G}_O(\omega)$. In addition, with the help of the inverse Fourier transform
\begin{equation}
\label{eq:17}
G_O(t) = \int_\mathbb{R} e^{i\omega t} \bar{G}_O(\omega)d\omega,
\end{equation}
one also proves that
\[
D_{2k} = (-1)^k \frac{d^{2k}G_O}{dt^{2k}}(0) = \int_{\mathbb{R}} \bar{G}_O(\omega)\omega^{2k}d\omega.
\]

We summarize the findings of the present subsection into the following proposition.
\begin{2}
\label{Pr:2}
The prescription
\begin{equation}
\label{eq:18}
dP_O(\omega)= \frac{1}{D_0} \bar{G}_O(\omega) d\omega
\end{equation}
defines a symmetric probability measure on $\mathbb{R}$. Thus, the odd moments $\mu_{2k+1}$ of the measure are zero. The even moments of the probability measure $dP_O(\omega)$ are finite and equal to
\begin{equation}
\label{eq:19}
\mu_{2k} \equiv \int_{\mathbb{R}}\omega^{2k}dP_O(\omega) = \frac{D_{2k}}{D_0}, \quad \forall k \geq 1.
\end{equation}
\end{2}

\subsection{The moment problem to be solved}

Surely, the reader has already anticipated that the problem we want to solve is the following Hamburger moment problem: \emph{Determine the symmetric probability measure $dP_O(\omega)$  on $\mathbb{R}$, the even moments of which are given by the sequence  $\{D_{2k}/{D_0},\; k\geq 1\}$.} However, in order for the problem to be correctly formulated, we must show that there exists a unique symmetric probability measure of even moments $\{D_{2k}/{D_0},\; k\geq 1\}$. 

The existence is automatically guarantied by the prescription $ [\bar{G}_O(\omega)/D_O] d\omega$, the normalized physical spectral density, which furnishes an example. For uniqueness, we cite the following theorem (Th.~3.11 from Section~2.3 of Ref.~\onlinecite{Dur96}).
\begin{3}
\label{Th:1}
If $\limsup_{k \to \infty} \mu_{2k}^{1/2k}/2k  < \infty$, then there is at most one distribution function $P_O(\omega)$ with $\mu_k = \int \omega^k dP_O(\omega)$ for all positive integers $k$.
\end{3}

We then have the following theorem.
\begin{4}
\label{Th:2}
There exists a unique symmetric probability measure $dP_O(\omega)$ of even moments $\{D_{2k}/{D_0},\; k\geq 1\}$, which is the one associated with the physical spectral weight function. Consequently, the sequence of positive numbers $\{D_{2k},\; k\geq 0\}$ uniquely determines the autocorrelation function $G_O(t)$ on the whole real axis.
\end{4}

\emph{Proof.} Let $t = \hbar \beta / 4$ and $a = G_0(it)$. From Eq.~(\ref{eq:14}) we learn that $D_{2k} \leq a {(2k)!}/{t^{2k}}$. With the help of Stirling's formula, we compute
\begin{eqnarray}
\label{eq:20} \nonumber 
\limsup_{k \to \infty} \frac{1}{2k} \left(\frac{ D_{2k}}{D_0}\right)^{1/2k} \leq \frac{1}{t} \lim_{k \to \infty}  \left(\frac{a}{D_0}\right)^{1/2k} \quad \\ \times \frac{[(2k)!]^{1/2k}}{2k}   =  \frac{1}{t} \lim_{k \to \infty} \frac{1}{2k}  \left[\frac{(2k)^{2k}\sqrt{4\pi k}}{e^{2k}}\right]^{1/2k} \\ = \frac{1}{e\cdot t}\lim_{k \to \infty} (4\pi k)^{1/4k} = \frac{1}{e\cdot t} < \infty \nonumber
\end{eqnarray}
and the theorem follows from Th.~\ref{Th:1} and the uniqueness of the inverse Fourier transforms of probability distributions (so-called \emph{characteristic functions} of the respective probability measures, according to Section~2.3.a of Ref.~\onlinecite{Dur96}). \hspace{\stretch{1}}$\Box$

In particular, Th.~\ref{Th:2} shows that the dynamics on the whole line is in principle uniquely determined by the sequence of derivatives at origin of the imaginary-time correlation function. Of course, this also follows from Baym and Mermin's analytic continuation result, but the proof we have performed is more direct in the sense that it connects the uniqueness with the numerical technique in a straightforward fashion. The reader will appreciate this from the following theorem, which gives general criteria for the pointwise recovery of the correlation function $G_O(t)$ on the whole real axis. 

\begin{5}
\label{Th:3}
Let $dP_{O,n}(\omega)$ be a sequence of symmetric probability measures such that 
\[\lim_{n \to \infty}\int_\mathbb{R}\omega^{2k} dP_{O,n}(\omega) = D_{2k}/D_0\] for each $k \geq 1$. Then 
\[\lim_{n \to \infty} G_{O,n}(t) = G_{O}(t), \; \forall t \in \mathbb{R}.\]
\end{5}

\emph{Observation.} Of course, by $G_{O,n}(t)$ we understand, up to a multiplication factor of $D_0$, the characteristic function of the measure $dP_{O,n}(\omega)$. The characteristic function is defined by
\[G_{O,n}(t) = D_0 \int_\mathbb{R} e^{i\omega t} dP_{O,n}(\omega).
\]
Remembering Eqs.~(\ref{eq:17}) and (\ref{eq:18}), we see that $G_O(t)$ is also a characteristic function, namely that of the measure $dP_O(\omega)$, because
\[G_{O}(t) = D_0 \int_\mathbb{R} e^{i\omega t} dP_{O}(\omega).
\]
Characteristic functions of measures are always \emph{continuous}, fact that follows easily from the dominated convergence theorem. 

\emph{Proof of Th.~\ref{Th:3}.} Th.~3.12 from Section~2.3 of Ref.~\onlinecite{Dur96} asserts that the sequence of probability measures $dP_{O,n}(\omega)$ converges weakly to $dP_{O}(\omega)$, because Eq.~(\ref{eq:20}) holds true. The first part of the continuity theorem (Th.~3.4 from Section~2.3 of the same reference) states that the weak convergence of the probability measures implies pointwise convergence of the corresponding characteristic functions at all times $t \in \mathbb{R}$. The last observation concludes the proof of the theorem. \hspace{\stretch{1}}$\Box$

In a sense, Th.~\ref{Th:3} says that the pointwise values of the correlation functions are the easiest to obtain. Basically, any procedure that is capable of reproducing the first $n$ moments of the true probability distribution leads to convergence of the correlation functions, in the limit of large $n$. Other properties, as for instance certain integral values involving correlation functions, are more difficult to obtain. Given the general approach put forward in the present section, we are now ready to discuss the two main computational aspects of the technique: the computation of the sequence of even derivatives of the imaginary-time correlation function and the numerical resolution of the associated Hamburger moment problem.

\section{Derivatives of the imaginary-time correlation functions}

According to Proposition~1, the Taylor series about origin of the imaginary-time correlation function $G_O(it)$ is convergent in the disk of equation $|t| < \beta \hbar / 2$ of the complex plane. As the well-known example of the free particle flux autocorrelation function (see Eq.~\ref{eq:54}) demonstrates, in general, one cannot expect convergence beyond this radius. Thus, for the purpose of computing derivatives in origin of the imaginary-time correlation function, we are forced to restrict the range of values of $t$ on which $G_O(it)$ is ``sampled'' to  the real interval $(-\beta \hbar/2, \beta \hbar / 2)$. On this interval, the correlation function $G_O(it)$  is computable with the help of the Feynman-Kac formula\cite{Fey48, Pre02, Sim79}  and we now turn our attention to the problem of constructing path-integral estimators for the evaluation of the high-order derivatives of  $G_O(it)$. 

We shall illustrate the general strategy for the derivation of  estimators for the particular case of the flux autocorrelation function. The reader needs notice that, following the prescription of Predescu and Doll,\cite{Pre02}  we strive to bury the time dependence into the potential part of the various estimators in order for these estimators to have finite variance in the limit of an infinite number of path variables. This  procedure prevents the well-known divergence of the variances of the estimators obtained by direct differentiation against imaginary time, with the increase of the number of path variables. Such a divergence is characteristic of the Barker estimators\cite{Bar79, Her82} and  is caused by an unfortunate attempt to differentiate the Brownian paths entering the Feynman-Kac formula (a famous 1933 theorem of Paley, Wiener, and Zygmund says that Brownian paths are not differentiable, with probability one).\cite{Pal33}  In addition, at the cost of utilizing a one-dimensional finite-difference scheme, the approach avoids the computation of the high order derivatives of the potential that appear in virial estimators\cite{Her82} as well as in estimators for which the imaginary-time differentiation is replaced by the direct action of the Hamiltonian. Even more, available numerical results (it is true, for low order derivatives, only) suggest that the variances of  thermodynamic estimators we utilize are smaller than the variances for the corresponding virial\cite{Pre03} and Hamiltonian techniques,\cite{Pre03c} especially at low temperature.

For a one-dimensional system, the imaginary-time flux autocorrelation function reads\cite{Mil83, Mil98}
\begin{equation}
\label{eq:21}
G_{F}(it)= \mathrm{tr}\left(e^{-(\beta/2 + t/\hbar) H}\hat{F} e^{-(\beta/2 - t/\hbar) H} \hat{F}\right),
\end{equation}
where 
\begin{equation}
\label{eq:22}
\hat{F} = \frac{1}{2m_0}\left[\delta(\hat{x}-x_s)\hat{p} + \hat{p}\, \delta\left(\hat{x}-x_s\right)\right]
\end{equation}
and
\[
\hat{p} = \frac{\hbar}{i}\frac{\partial}{\partial x}
\]
are self-adjoint operators (therefore, $\hat{F}^\dagger = \hat{F}$). The flux operator $\hat{F}$ corresponds to the dividing surface passing through $x_s$ (actually, a ``dividing point'' in this one-dimensional case). Setting 
$\beta_t = \beta /2 + t/\hbar$, Eq.~(\ref{eq:21}) takes the form
\begin{eqnarray}
\label{eq:23}\nonumber
G_{F}(it) &=& \left(\frac{\hbar}{2m_0}\right)^2 \bigg[ \rho\left(x,x';\beta_{-t}\right)  \frac{\partial^2\rho}{\partial x\partial x'} \left(x,x';\beta_{t}\right)  \\ &&
\nonumber +  \frac{\partial^2 \rho}{\partial x\partial x'} \left(x,x';\beta_{-t}\right) \rho\left(x,x';\beta_{t}\right)    \\ &&
 -  \frac{\partial \rho}{\partial x} \left(x,x';\beta_{-t}\right)   \frac{\partial \rho}{\partial x'} \left(x,x';\beta_{t}\right)   \\ &&
 \nonumber -  \frac{\partial \rho}{\partial x'} \left(x,x';\beta_{-t}\right) \frac{\partial \rho}{\partial x}\left(x,x';\beta_{t}\right)  \bigg] \bigg|_{x' = x = x_s}, 
\end{eqnarray}
where, of course, $\rho(x,x';\beta_t)$ is the density matrix at the inverse temperature $\beta_t$.

Let us consider the one-dimensional Feynman-Kac formula\cite{Fey48, Pre02, Sim79} 
\begin{equation}
\label{eq:24}
\rho(x,x';\beta_t)= \rho_{fp}(x,x';\beta_t) \mathbb{E} e^{-\beta_t \int_0^1V\left[x_r(u)+ \sigma_t B_u^0\right]du},
\end{equation} 
which expresses the density matrix as the expected value of a functional of the standard Brownian bridge $B_u^0$. In Eq.~(\ref{eq:24}), $x_r(u) = x + (x' - x)u$ and $\sigma_t = (\hbar^2\beta_t/m_0)^{1/2}$, whereas $\rho_{fp}(x,x';\beta_t)$ stands for the density matrix of a similar free particle at the inverse temperature $\beta_t$. 
By explicit computation, from Eq.~(\ref{eq:23}) and the Feynman-Kac formula, one derives the equation
\begin{eqnarray}
\label{eq:25}\nonumber
G_{F}(it) =  \mathbb{E}\mathbb{E}' e^{-\beta_{-t} \int_0^1V\left(x_s+ \sigma_{-t} B_u^0\right)du-\beta_t \int_0^1V\left(x_s+ \sigma_t {B_u^0}'\right)du}  \\ \times  \left(\frac{\hbar}{2m_0}\right)^2 \rho_{fp}(0;\beta_{-t})\rho_{fp}(0;\beta_t)\mathcal{F}_t^0\left(B_\star^0, {B_\star^0}'\right), \quad
\end{eqnarray}
where
\begin{widetext}
\begin{eqnarray}
\label{eq:26}
\nonumber
\mathcal{F}_t^0\left(B_\star^0, {B_\star^0}'\right) &=&  \frac{1}{2 \sigma_{-t}^2} + \frac{1}{2 \sigma_t^2} + \beta_{t}^2 \left[\int_0^1V'\left(x_s+ \sigma_{t} {B_u^0}'\right)udu\right]\left[\int_0^1V'\left(x_s+ \sigma_{t} {B_u^0}'\right)(1-u)du\right] 
\nonumber \\ && +  \beta_{-t}^2 \left[\int_0^1V'\left(x_s+ \sigma_{-t} {B_u^0}\right)udu\right] \left[\int_0^1V'\left(x_s+ \sigma_{-t} {B_u^0}\right)(1-u)du\right] 
\nonumber \\ && - \beta_{-t} \beta_t \left[\int_0^1V'\left(x_s+ \sigma_{-t} B_u^0\right)udu\right]\left[\int_0^1V'\left(x_s+ \sigma_{t} {B_u^0}'\right)(1-u)du\right]
 \\ && - \beta_{-t} \beta_t \left[\int_0^1V'\left(x_s+ \sigma_{t} {B_u^0}'\right)udu\right]\left[\int_0^1V'\left(x_s+ \sigma_{-t} {B_u^0}\right)(1-u)du\right]
 \nonumber \\ && \nonumber
 - \beta_{-t}\int_0^1V''\left(x_s + \sigma_{-t} B_u^0\right)u(1-u)du - \beta_{t}\int_0^1V''\left(x_s+ \sigma_{t} {B_u^0}'\right)u(1-u)du.
\end{eqnarray}
\end{widetext}
In Eq.~(\ref{eq:25}), the symbols $\mathbb{E}$ and $\mathbb{E}'$ denote the expected values against the independent  standard Brownian bridges $B_u^0$ and ${B_u^0}'$, respectively. In Eq.~(\ref{eq:26}), $V'(x)$ and $V''(x)$ denote the first and the second derivatives of the potential $V(x)$, respectively. 

Now, Eq.~(\ref{eq:25}) can be rearranged as 
\begin{eqnarray}
\label{eq:27}\nonumber
G_{F}(it) =  \mathbb{E}\mathbb{E}' e^{-(\beta/2) \left[\int_0^1V\left(x_s+ \sigma_0 B_u^0\right)du+\int_0^1V\left(x_s+ \sigma_0 {B_u^0}'\right)du\right]}  \\ \times \frac{1}{8\pi m_0}\mathcal{F}'_t\left(B_\star^0, {B_\star^0}'\right), \quad 
\end{eqnarray}
where
\begin{eqnarray}
\label{eq:28}
 \nonumber 
\mathcal{F}'_t\left(B_\star^0, {B_\star^0}'\right)= \frac{1}{\sqrt{\beta_{-t}\beta_t}}  \mathcal{F}_t^0\left(B_\star^0, {B_\star^0}'\right) \\ \times  e^{-(\beta/2)\left[ \Delta_{-t}\left(B_\star^0\right) + \Delta_t\left({B_\star^0}'\right)\right]}, 
\end{eqnarray}
and
\begin{eqnarray}
\label{eq:29} \nonumber
 \Delta_{t}\left(B_\star^0\right) = \int_0^1V\left(x_s+ \sigma_0 B_u^0\right)du \\ - \frac{2\beta_{t}}{\beta} \int_0^1V\left(x_s+ \sigma_{t} B_u^0\right)du.
\end{eqnarray}

Anticipating the use of Monte Carlo techniques for the evaluation of imaginary-time correlation functions and related properties, we introduce the normalization factor
\begin{equation}
\label{eq:30} 
\mathcal{N}_F=\frac{1}{8\pi m_0} \mathbb{E}\mathbb{E}' e^{-(\beta/2) \left[\int_0^1V\left(x_s+ \sigma_0 B_u^0\right)du+\int_0^1V\left(x_s+ \sigma_0 {B_u^0}'\right)du\right]}. 
\end{equation}
In principle, the factor $\mathcal{N}_F$ can be evaluated in a separate Monte Carlo simulation, although for the one-dimensional example presented later in the paper, we shall employ the numerical matrix multiplication technique.\cite{Kle73, Thi83a} If rate constants rather than absolute rates of reaction are desired, one seeks to evaluate the ratio between $\mathcal{N}_F$ and the  partition function of the reactant side, $Q_r$. A Monte Carlo approach to the computation of such ratios has been recently presented in Ref.~\onlinecite{Yam04}. 

In any case, the main difficulty in the computation of quantum correlation functions does not reside in the evaluation of the normalization coefficient $\mathcal{N}_F$. Therefore, for the remainder of the present paper, we shall focus our attention on the Monte Carlo evaluation of the ratios 
\begin{widetext}
\begin{eqnarray}
\label{eq:31}
\frac{G_F(it)}{\mathcal{N}_F}= \left\langle \mathcal{F}'_t\left(B_\star^0, {B_\star^0}'\right)  \right \rangle = \frac{\mathbb{E}\mathbb{E}' e^{-(\beta/2) \left[\int_0^1V\left(x_s+ \sigma_0 B_u^0\right)du+\int_0^1V\left(x_s+ \sigma_0 {B_u^0}'\right)du\right]} \mathcal{F}'_t\left(B_\star^0, {B_\star^0}'\right)}{\mathbb{E}\mathbb{E}' e^{-(\beta/2) \left[\int_0^1V\left(x_s+ \sigma_0 B_u^0\right)du+\int_0^1V\left(x_s+ \sigma_0 {B_u^0}'\right)du\right]} },
\end{eqnarray}
\end{widetext}
or related quantities.  For the purpose of computing averages of the type given by Eq.~(\ref{eq:31}), it turns out that it is useful to replace the estimating function $\mathcal{F}'_t\left(B_\star^0, {B_\star^0}'\right)$ with the symmetric form
\begin{equation}
\label{eq:32}
\mathcal{F}_t\left(B_\star^0, {B_\star^0}'\right) = \frac{1}{2}\left[\mathcal{F}'_{-t}\left(B_\star^0, {B_\star^0}'\right) + \mathcal{F}'_t\left(B_\star^0, {B_\star^0}'\right) \right]. 
\end{equation}
As follows from the equation $G_F(-it) = G_F(it)$, this replacement does not change the value of  $G_F(it)$. However, in the next paragraph, we shall prove that  the resulting estimator has a smaller variance.

It follows from Eqs.~(\ref{eq:26}) and (\ref{eq:28}) that
\begin{equation}
\label{eq:32a}
\mathcal{F}'_{-t}\left(B_\star^0, {B_\star^0}'\right) = \mathcal{F}'_t\left({B_\star^0}', B_\star^0\right)
\end{equation}  
and therefore, 
\begin{eqnarray}
\label{eq:33} \nonumber
\mathcal{F}_t\left(B_\star^0, {B_\star^0}'\right) = \frac{1}{2}\left[\mathcal{F}'_t\left({B_\star^0}', B_\star^0\right) \right. \\ \left. + \mathcal{F}'_t\left(B_\star^0, {B_\star^0}'\right) \right]  = \mathcal{F}_t\left({B_\star^0}', B_\star^0\right). 
\end{eqnarray}
Consequently, the function $\mathcal{F}_t\left(B_\star^0, {B_\star^0}'\right)$ is not only symmetric with respect to time inversion, as follows directly from Eq.~(\ref{eq:31}), but also with respect to the exchange of variables $B_\star^0$ and ${B_\star^0}'$. 
Let us write $\mathcal{F}'_t\left(B_\star^0, {B_\star^0}'\right)$ as the sum between its symmetric and its antisymmetric parts
\begin{eqnarray*} &&
\mathcal{F}'_t\left(B_\star^0, {B_\star^0}'\right) = \mathcal{F}_t\left(B_\star^0, {B_\star^0}'\right)  \\ && + \frac{1}{2}\left[\mathcal{F}'_{t}\left(B_\star^0, {B_\star^0}'\right) - \mathcal{F}'_{t}\left({B_\star^0}', B_\star^0\right) \right].
\end{eqnarray*}
Since antisymmetric functions integrate to zero against a symmetric probability measure, and since the products of  symmetric and antisymmetric functions are antisymmetric, we have
\begin{eqnarray*} &&
\left\langle \mathcal{F}'_t\left(B_\star^0, {B_\star^0}'\right)^2 \right\rangle = \left\langle \mathcal{F}_t\left(B_\star^0, {B_\star^0}'\right)^2 \right\rangle \\ && + \frac{1}{4} \left\langle  \left[\mathcal{F}'_{t}\left(B_\star^0, {B_\star^0}'\right) - \mathcal{F}'_{t}\left({B_\star^0}', B_\star^0\right) \right]^2\right\rangle.
\end{eqnarray*}
The last equation and the equality 
\[
\left\langle \mathcal{F}'_t\left(B_\star^0, {B_\star^0}'\right) \right\rangle = \left\langle \mathcal{F}_t\left(B_\star^0, {B_\star^0}'\right) \right\rangle = \frac{G_F(it)}{G_d(0)},
\]
which was discussed in the previous paragraph, clearly demonstrate that the estimator given by Eq.~(\ref{eq:32}) has a  variance smaller than that of the estimator given by Eq.~(\ref{eq:28}). 

To summarize, by Monte Carlo simulations, one may compute averages of the type
\begin{widetext}
\begin{eqnarray}
\label{eq:34}
\frac{G_F(it)}{\mathcal{N}_F}= \left\langle \mathcal{F}_t\left(B_\star^0, {B_\star^0}'\right)  \right \rangle = \frac{\mathbb{E}\mathbb{E}' e^{-(\beta/2) \left[\int_0^1V\left(x_s+ \sigma_0 B_u^0\right)du+\int_0^1V\left(x_s+ \sigma_0 {B_u^0}'\right)du\right]} \mathcal{F}_t\left(B_\star^0, {B_\star^0}'\right)}{\mathbb{E}\mathbb{E}' e^{-(\beta/2) \left[\int_0^1V\left(x_s+ \sigma_0 B_u^0\right)du+\int_0^1V\left(x_s+ \sigma_0 {B_u^0}'\right)du\right]} },
\end{eqnarray}
\end{widetext}
where 
\begin{eqnarray}
\label{eq:35}
 \nonumber 
\mathcal{F}_t\left(B_\star^0, {B_\star^0}'\right)= \frac{1}{2\sqrt{\beta_{-t}\beta_t}}   \left\{\mathcal{F}_{-t}^0\left(B_\star^0, {B_\star^0}'\right) \right. \\ \times  e^{-(\beta/2)\left[ \Delta_{t}\left(B_\star^0\right) + \Delta_{-t}\left({B_\star^0}'\right)\right]} + \mathcal{F}_t^0\left(B_\star^0, {B_\star^0}'\right) \\ \left.
\times   e^{-(\beta/2)\left[ \Delta_{-t}\left(B_\star^0\right) + \Delta_t\left({B_\star^0}'\right)\right]}\right\}. \nonumber
\end{eqnarray}
The estimating function $\mathcal{F}_t\left(B_\star^0, {B_\star^0}'\right)$ is symmetric under time inversion  --- that is, $\mathcal{F}_t\left(B_\star^0, {B_\star^0}'\right) = \mathcal{F}_{-t}\left(B_\star^0, {B_\star^0}'\right)$ --- as well as under the exchange of the variables $B_\star^0$ and ${B_\star^0}'$.

The construction of  estimators for derivatives in origin is straightforward and follows from Eq.~(\ref{eq:34}). By Monte Carlo simulations, one may compute the following averages
\begin{widetext}
\begin{equation}
\label{eq:36}
\frac{1}{\mathcal{N}_F}\left. \frac{d^k}{dt^k}G_F(it)\right|_{t=0}= \frac{\mathbb{E}\mathbb{E}' e^{-(\beta/2) \left[\int_0^1V\left(x_s+ \sigma B_u^0\right)du+\int_0^1V\left(x_s+ \sigma {B_u^0}'\right)du\right]} \frac{d^k}{dt^k}\mathcal{F}_t\left(B_\star^0, {B_\star^0}'\right)\Big|_{t = 0}}{\mathbb{E}\mathbb{E}' e^{-(\beta/2) \left[\int_0^1V\left(x_s+ \sigma B_u^0\right)du+\int_0^1V\left(x_s+ \sigma {B_u^0}'\right)du\right]} }.
\end{equation}
\end{widetext}
In this respect, the reader should notice that the function $\mathcal{F}_t\left(B_\star^0, {B_\star^0}'\right)$ is well-defined for all $t \in (-\beta \hbar / 2, \beta \hbar / 2)$ and is infinitely differentiable on this interval provided that the potential $V(x)$ is also differentiable infinitely many times. In practical applications, the time derivatives appearing in Eq.~(\ref{eq:36}) are to be computed by finite difference. We shall further discuss this matter in Section~IV.

We now describe the construction of estimators for the case of a $d$-dimensional system. For definiteness, we shall assume that the physical coordinates have been rescaled such that all masses are equal to the common value $m_0$. Perhaps after a reorientation of the system of axes so that the first coordinate $\mathbf{x}_1$ is along the reaction coordinate, the reactants and products are assumed to be separated in the configuration space $\mathbb{R}^d$ by a hyperplane of equation $\mathbf{x}_1 = x_s$. For the remainder of this section, when dealing with expressions involving the density matrix,  it turns out that it is more convenient to work with the pair of position coordinates $(\mathbf{x}, \mathbf{z})$, with $\mathbf{z} = \mathbf{x}' - \mathbf{x}$, rather than with the standard $(\mathbf{x}, \mathbf{x}')$ pair. This is so because identities of the type
\begin{eqnarray}
\label{eq:37} \nonumber
\int_\mathbb{R} dx' \rho_{fp}(x,x';\beta_t)\rho_{fp}(x,x';\beta_{-t})f( x' - x) \\ = 
\frac{1}{2\pi \sigma_0} \int_\mathbb{R} dz e^{-z^2} f\left(z\sigma_{\pm t}\right),
\end{eqnarray}
where $\sigma_{\pm t} =\sigma_t\sigma_{-t}/\sigma_0 $, are clearly simpler to express in the new coordinate system. Moreover, transformations of the type shown by Eq.~(\ref{eq:37}) are consistent with the aforementioned advice of Predescu and Doll that the time dependence of paths should be buried into the potential part of the Feynman-Kac formula whenever possible. 

With these clarifications, we leave it for the reader to demonstrate that the multidimensional  analogues of the various quantities necessary for the construction of derivative estimators are as follows. With the understanding that the quantities $V'(\mathbf{x})$ and $V''(\mathbf{x})$ now denote the first order and the second order partial derivatives against the reaction coordinate $\mathbf{x}_1$, the multi-dimensional analogue of Eq.~(\ref{eq:26}) is
\begin{widetext}
\begin{eqnarray}
\label{eq:38}
\nonumber
\mathcal{F}_t^0\left(\mathbf{x}, \mathbf{z}, B_\star^0, {B_\star^0}'\right) &=&  \frac{1}{2 \sigma_{-t,0}^2} + \frac{1}{2 \sigma_{t,0}^2} 
\nonumber \\ && + \beta_{t}^2 \left[\int_0^1V'\left(\mathbf{x} + \sigma_{\pm t} \mathbf{z} u+ \sigma_{t} {B_u^0}'\right)udu\right]\left[\int_0^1V'\left(\mathbf{x} + \sigma_{\pm t} \mathbf{z} u + \sigma_{t} {B_u^0}'\right)(1-u)du\right] 
\nonumber \\ && +  \beta_{-t}^2 \left[\int_0^1V'\left(\mathbf{x} + \sigma_{\pm t} \mathbf{z} u+ \sigma_{-t} {B_u^0}\right)udu\right] \left[\int_0^1V'\left(\mathbf{x} + \sigma_{\pm t} \mathbf{z} u+ \sigma_{-t} {B_u^0}\right)(1-u)du\right] 
\nonumber \\ && - \beta_{-t} \beta_t \left[\int_0^1V'\left(\mathbf{x} + \sigma_{\pm t} \mathbf{z} u+ \sigma_{-t} B_u^0\right)udu\right]\left[\int_0^1V'\left(\mathbf{x} + \sigma_{\pm t} \mathbf{z} u+ \sigma_{t} {B_u^0}'\right)(1-u)du\right]
 \\ && - \beta_{-t} \beta_t \left[\int_0^1V'\left(\mathbf{x} + \sigma_{\pm t} \mathbf{z} u+ \sigma_{t} {B_u^0}'\right)udu\right]\left[\int_0^1V'\left(\mathbf{x} + \sigma_{\pm t} \mathbf{z} u+ \sigma_{-t} {B_u^0}\right)(1-u)du\right]
 \nonumber \\ && \nonumber
 - \beta_{-t}\int_0^1V''\left(\mathbf{x} + \sigma_{\pm t} \mathbf{z} u + \sigma_{-t} B_u^0\right)u(1-u)du - \beta_{t}\int_0^1V''\left(\mathbf{x} + \sigma_{\pm t} \mathbf{z} u+ \sigma_{t} {B_u^0}'\right)u(1-u)du.
\end{eqnarray}
\end{widetext}
The quantities $B_u^0$ and ${B_u^0}'$ are independent $d$-dimensional standard Brownian bridges ($d$-dimensional vector valued stochastic processes, the components of which are independent one-dimensional standard Brownian bridges).
We also define
\begin{eqnarray}
\label{eq:39} \nonumber
 \Delta_{t}\left(\mathbf{x}, \mathbf{z}, B_\star^0\right) = \int_0^1V\left(\mathbf{x} + \sigma_{0} \mathbf{z} u + \sigma_0 B_u^0\right)du \\ - \frac{2\beta_{t}}{\beta} \int_0^1V\left(\mathbf{x} + \sigma_{\pm t} \mathbf{z} u + \sigma_{t} B_u^0\right)du
\end{eqnarray}
as well as
\begin{eqnarray}
\label{eq:40}
 \nonumber 
\mathcal{F}_t\left(\mathbf{x}, \mathbf{z}, B_\star^0, {B_\star^0}'\right)= \frac{1}{2\sqrt{\beta_{-t}\beta_t}}   \left\{\mathcal{F}_{-t}^0\left(\mathbf{x}, \mathbf{z}, B_\star^0, {B_\star^0}'\right) \right. \\ \times  e^{-(\beta/2)\left[ \Delta_{t}\left(\mathbf{x}, \mathbf{z}, B_\star^0\right) + \Delta_{-t}\left(\mathbf{x}, \mathbf{z}, {B_\star^0}'\right)\right]} + \mathcal{F}_t^0\left(\mathbf{x}, \mathbf{z}, B_\star^0, {B_\star^0}'\right) \quad \\ \left. \times   e^{-(\beta/2)\left[ \Delta_{-t}\left(\mathbf{x}, \mathbf{z}, B_\star^0\right) + \Delta_t\left(\mathbf{x},\mathbf{z},{B_\star^0}'\right)\right]}\right\}. \nonumber
\end{eqnarray}
The normalization coefficient $\mathcal{N}_F$ now reads 
\begin{widetext}
\begin{equation}
\label{eq:41}
\mathcal{N}_F = \frac{1}{8\pi m_{0}}\left(\frac{1}{2\pi \sigma_{0}}\right)^{d-1}\int_{\mathcal{S}}d \mathbf{x} d\mathbf{z}\mathbb{E}\mathbb{E}'e^{-\|\mathbf{z}\|^2} e^{-(\beta/2) \left[\int_0^1V\left(\mathbf{x} + \sigma_0 \mathbf{z} u + \sigma_0 B_u^0\right)du+\int_0^1V\left(\mathbf{x} + \sigma_0 \mathbf{z} u + \sigma_0 {B_u^0}'\right)du\right]}, 
\end{equation}
\end{widetext}
where the integration against the variables $\mathbf{x}$ and $\mathbf{z}$ is done on the $(d - 2)$-dimensional hyperplane $\mathcal{S}$, which is the subset of the space $\mathbb{R}^d \times \mathbb{R}^d$ defined by the equations ${x}_1 = x_s$ and ${z}_1 = 0$. Therefore, the symbol $d\mathbf{x}$ stands for the Lebesgue measure $d{x}_2  \cdots d{x}_d$, whereas $d\mathbf{z}$ stands for  $d{z}_2  \cdots d{z}_d$. The Euclidian norm $\|\mathbf{z}\| = ({z}_1^2 + \cdots + {z}_d^2)^{1/2}$ can be replaced by $\|\mathbf{z}\| = ({z}_2^2 + \cdots + {z}_d^2)^{1/2}$, since the coordinate ${z}_1$ is kept constant and equal to zero during integration. 

In these conditions, up to the value of the normalization coefficient $\mathcal{N}_F$, the derivatives in origin of the flux autocorrelation functions can be determined by Monte Carlo integration, as implied by the equation
\begin{widetext}
\begin{eqnarray}\nonumber
\label{eq:42}&&
\frac{D_k}{\mathcal{N}_F} = \frac{1}{\mathcal{N}_F}\left. \frac{d^k}{dt^k}G_F(it)\right|_{t=0} \\ && =  \frac{\int_{\mathcal{S}}d \mathbf{x} d\mathbf{z}\mathbb{E}\mathbb{E}' e^{-\|\mathbf{z}\|^2}e^{-(\beta/2) \left[\int_0^1V\left(\mathbf{x} + \sigma_0 \mathbf{z} u + \sigma_0 B_u^0\right)du+\int_0^1V\left(\mathbf{x} + \sigma_0 \mathbf{z} u + \sigma_0 {B_u^0}'\right)du\right]} \frac{d^k}{dt^k}\mathcal{F}_t\left(\mathbf{x}, \mathbf{z}, B_\star^0, {B_\star^0}'\right)\Big|_{t = 0}}{\int_{\mathcal{S}}d \mathbf{x} d\mathbf{z}\mathbb{E}\mathbb{E}' e^{-\|\mathbf{z}\|^2} e^{-(\beta/2) \left[\int_0^1V\left(\mathbf{x} + \sigma_0 \mathbf{z} u + \sigma_0 B_u^0\right)du+\int_0^1V\left(\mathbf{x} + \sigma_0 \mathbf{z} u + \sigma_0 {B_u^0}'\right)du\right]} }.
\end{eqnarray}
\end{widetext}

\section{Solving the inverse moment problem: a numerical example}

Until now, we have demonstrated that the sequence of derivatives at origin completely and uniquely characterizes the correlation function. Moreover, the sequence of derivatives can be computed by Monte Carlo simulation via estimators that have finite variance in the limit of an infinite number of path variables (of course, for analytic potentials). At this point, it is natural to address the problem of recovering the correlation functions  from the sequence of computed moments. 

More precisely, let us assume that we have computed the set of even and nonnegative derivatives $D_0, D_2, \ldots, D_{2n}$ and that we have calculated the moments $\mu_{2k} = D_{2k}/D_0$, for $1 \leq k \leq n$.  At the very least, we would like to construct a sequence of  symmetric probability distributions $dP_{O,n}(\omega)$ such that 
\begin{equation}
\label{eq:43}
\mu_{2k}=\int_{\mathbb{R}}\omega^{2k} dP_{O,n}(\omega),
\end{equation}
for all $1 \leq k \leq n$ and $n \geq 1$. Indeed, if Eq.~(\ref{eq:43}) is satisfied, then so is the hypothesis of Th.~\ref{Th:3}, theorem that further guaranties that the correlation functions are fully recovered (pointwise) in the limit $n \to \infty$. However, many times, the pointwise reconstruction of the correlation functions does not suffice. For example, in the case of the flux autocorrelation function, the chemical physicists are usually interested in computing the absolute rate of reaction, which is the time integral of the correlation function
\begin{equation}
\label{eq:44}
k(T)Q_r(T) = \int_0^\infty G_F(t)dt.
\end{equation}
Because the first $n$ even moments  do not uniquely determine a symmetric probability distribution, we have freedom in choosing the reconstruction algorithm in such a way that not only the pointwise values of the correlation functions, but also various integral expressions are recovered in the limit $n \to \infty$.

Although the optimal reconstruction algorithm depends upon the nature of the correlation functions and of the quantum information being sought, we shall discuss and utilize in the present paper a choice that is based on the maximum entropy approach. The maximum entropy method\cite{Jay57, Jay78, Ski89a, Gul89a, Jar96} suggests that a useful criterion is to chose the probability distribution $\bar{G}(\omega)$ that maximizes the Shanon entropy
\begin{equation}
\label{eq:a0}
S(\bar{G})= - \int_\mathbb{R} \bar{G}(\omega)\ln\left[\bar{G}(\omega)/m(\omega)\right]d\omega,
\end{equation}
relative to the default model $m(\omega)$ and subject to the constraints
\begin{equation}
\label{eq:a1}
\int_\mathbb{R}\bar{G}(\omega)\omega^{2k}d\omega = D_{2k}, \quad 0 \leq k \leq n.
\end{equation}
In information theory, such a probability distribution is the least biased one that is compatible with the partial information represented by the known first moments. The default model $m(\omega)$ is a strictly positive distribution. Although it has a definite probabilistic meaning only if it is integrable, non-integrable default models can also be used. The choice  $m(\omega) = 1$ is called the flat default model.  

Simple variational arguments and use of Lagrange multipliers show that the unique maximum of the above problem is realized for 
\begin{equation}
\label{eq:45}
\bar{G}_{O,n}(\omega) = m(\omega)\exp\left(- \sum_{j = 0}^n \lambda_j \omega^{2j} \right).
\end{equation}
The coefficients $\lambda_0, \ldots, \lambda_n$ are the Lagrange multipliers and can be determined from the equations
\begin{equation}
\label{eq:46}
D_{2k} = \int_{\mathbb{R}} m(\omega)\omega^{2k}\exp\left(- \sum_{j = 0}^n \lambda_j \omega^{2j}\right) d\omega, \quad 0 \leq k \leq n.
\end{equation} 
Notice that the form of the approximant given by Eq.~(\ref{eq:45}) ensures both the positivity and the symmetry of the power spectrum, properties that have been demonstrated in Section~II.
Then, the entropy of $\bar{G}_{O,n}(\omega)$ is given by 
\begin{eqnarray}
\label{eq:47} && \nonumber
S\left[\bar{G}_{O,n}\right] = - \int_\mathbb{R} \bar{G}_{O,n}(\omega) \\ &&
\quad \times \ln\left[\bar{G}_{O,n}(\omega)/m(\omega)\right] d\omega  = \sum_{j = 0}^n \lambda_j D_{2j}.
\end{eqnarray}

One of the advantages of the maximum entropy algorithm is that, by use of default models,  it may incorporate additional physical information that depends upon the nature of the quantum results being sought. However, for the present example, a flat default model has been utilized. Also, for the present application, the data have been assumed noiseless. The stability of the final results with respect to the errors in the input data has been found to be excellent, in part because the number of matched moments is small, but also because the different data are perfectly correlated (they are obtained in the same Monte Carlo run). Thus, the assumption of noiseless data is good. For larger numbers of included moments, more general approaches of Bayesian statistical inference with entropic priors also allow for the treatment of noise in the data, via likelihood functions.\cite{Jar96}

The system of equations (\ref{eq:46}) can be replaced by 
\begin{equation}
\label{eq:48}
\lambda_0 = \ln\left[\frac{1}{D_0}\int_\mathbb{R}m(\omega)e^{- \sum_{j = 1}^n \lambda_j \omega^{2j}} d\omega \right]
\end{equation}
and 
\begin{equation}
\label{eq:49}
D_{2k} = D_0 \frac{\int_\mathbb{R}m(\omega)\omega^{2k}e^{- \sum_{j = 1}^n \lambda_j \omega^{2j}} d\omega}{\int_\mathbb{R}m(\omega)e^{- \sum_{j = 1}^n \lambda_j \omega^{2j}} d\omega}, \quad 1 \leq k \leq n.
\end{equation}
It is then a simple exercise to verify that Eqs.~(\ref{eq:49}) are satisfied for all $1 \leq k \leq n$ provided that the $\lambda_j$'s represent the coordinates of the minimum of the entropy functional 
\begin{eqnarray}
\nonumber
\label{eq:50}
S\left[\bar{G}_{O,n}\right] = D_0 \ln\left[\frac{1}{D_0}\int_\mathbb{R}m(\omega)e^{- \sum_{j = 1}^n \lambda_j \omega^{2j}} d\omega\right] \\ + \sum_{j = 1}^n \lambda_j D_{2j}, \qquad
\end{eqnarray}
which is a convex function of $\lambda_1, \ldots, \lambda_n$. Due to the convexity of the function that is minimized, the minimum of Eq.~(\ref{eq:50}), if it exists, is unique. The necessary and sufficient conditions for the existence of the minimum are known in literature.\cite{Jay78, Tag98} In the present article, the minimization of Eq.~(\ref{eq:50}) has been carried out with the help of Newton's steepest descent technique. The Hessian matrix is evaluated explicitly and utilized to predict the direction along which to line-minimize. The Golden Section search is utilized to optimize along the computed direction. As discussed in Ref.~\onlinecite{Tag98}, the computation of the coefficients $\lambda_j$ becomes less and less stable as the number of matched moments increases and, depending upon the number of even derivatives considered, may require extended-precision arithmetics. 

In order to demonstrate its usefulness, we apply the moment technique to the problem of computing the quantum rate of reaction for a symmetric Eckart barrier at various temperatures. The parameters for the Eckart barrier are chosen to correspond approximately to the $\mathrm{H}+\mathrm{H}_2$ reaction.\cite{Mil03} The potential is
\begin{equation}
\label{eq:50a}
V(x) = V_0 \; \mathrm{sech}(ax)^2,
\end{equation} 
with the parameters $V_0 = 0.425~\mathrm{eV}$, $a = 1.36~\mathrm{a.u.}$, and $m_0 =  1060~\mathrm{a.u.}$ 

We evaluate the flux autocorrelation function and its first five even derivatives at origin by Monte Carlo simulations, as described in Section~III. The derivatives of the estimator $\mathcal{F}_t(B_\star^0, {B_\star^0}')$ appearing in Eq.~(\ref{eq:36}) are replaced by numerical approximations computed via central difference. Remembering that $\mathcal{F}_t(B_\star^0, {B_\star^0}')$ is symmetric under the transformation $t \mapsto -t$, the finite-difference formulas take on the general form
\begin{equation}
\label{eq:51}
\frac{d^{2k}}{dt^{2k}}\mathcal{F}_t(B_\star^0, {B_\star^0}') = \frac{1}{\tau^{2k}}\sum_{j = 0}^{5}c_{k,j} \mathcal{F}_{j \tau}(B_\star^0, {B_\star^0}') + \mathcal{O}(\tau^{12 -2k}), 
\end{equation}
where the coefficients $c_{j,k}$ are given in Table~\ref{Tab:I}. Numerical experiments  demonstrate that a time step of
\begin{equation}
\label{eq:52}
\tau = \frac{1}{64}\frac{\hbar \beta}{2}
\end{equation}
 is sufficient for a determination of the derivatives to an accuracy of less than $2\%$.

Regarding the computation of derivatives by finite difference, the range of values of $\tau$ that can be utilized depends on the order of the derivatives as well as on the numerical precision with which the computations are conducted. For the present paper, we employ the IEEE floating-point data type \emph{double} ($64$ bit) for the representation of real numbers. Increasing the order of the derivatives beyond $10$ requires use of extended-precision data types.\cite{Hid01}

\begingroup
\squeezetable
\begin{table}[!htbp]
\begin{tabular}{| c | c | c | c | c | c | c | } 
\hline
$2k$ & $c_{k, 0}$ & $c_{k,1}$ & $c_{k,2}$ & $c_{k, 3}$ & $c_{k, 4}$ & $c_{k, 5}$ \\
\hline
0 & 1 & 0 & 0 & 0 & 0 & 0 \\
\hline
2 & $-{5269}/{1800}$ & ${10}/{3}$ & $-{10}/{21}$ & ${5}/{63}$ & $-{5}/{504}$ & ${1}/{1575}$ \\
\hline
4 & ${1529}/{120}$ & $-{1669}/{90}$ & ${4369}/{630}$ & $-{541}/{420}$ & ${1261}/{7560}$ & $-{41}/{3780}$ \\
\hline
6 & $-{1023}/{20}$ & ${323}/{4}$ & $-39$ & ${87}/{8}$ & $-{19}/{12}$ & ${13}/{120}$ \\
\hline
8 & $154$ & $-252$ & $136$ & $-46$ & ${26}/{3}$ &  $-{2}/{3}$ \\
\hline
10 & $-252$ & $420$ & $-240$ & $90$ & $-20$ & $2$  \\
\hline
\end{tabular}
\caption{\label{Tab:I} Numerical values for the coefficients $c_{k,j}$ appearing in the finite-difference approximations of the derivatives of order $2k$.}
\end{table}
\endgroup

Many times, the chemical physicist takes the different approach of constructing models (and, therefore, empirical inversion techniques) that have already incorporated additional physical input.\cite{Han94} In such cases, the finite number of derivatives  that can be computed using the data type \emph{double} may suffice for many practical purposes. This is why it is appropriate to table the coefficients $c_{j,k}$, for the reader's convenience. General rules for computing derivatives of arbitrary orders and accuracy have been discussed elsewhere.\cite{Kha99} According to Eq.~(\ref{eq:51}), the accuracy of the finite-difference scheme is largest for the small-order derivatives and decreases for the larger-order derivatives, if all the information contained in the $6$ points at which $\mathcal{F}_t(B_\star^0, {B_\star^0}')$ is evaluated is to be taken into consideration. This is to our advantage, because the low-order derivatives are computed with increased precision despite the relatively large value of the discretization step $\tau$ demanded by the higher-order derivatives. 

For the sake of an example, in Table II, we present the Monte Carlo estimates of the first five even derivatives at origin for the Eckart barrier at the temperature of $100~\textrm{K}$. The derivatives have been evaluated in $10$ million Monte Carlo points with the help of the estimators introduced in Section~III. For the discretization of the Feynman-Kac formula, we employ Predescu's fourth-order path-integral technique\cite{Pre04} with a number of $64$ path variables. This technique is basically a Trotter product 
\begin{eqnarray} \nonumber 
\label{eq:II6}
\rho_n(x,x';\beta)=\int_{\mathbb{R}}d x_1 \ldots \int_{\mathbb{R}}d x_n\; \rho_0\left(x,x_1;\frac{\beta}{n+1}\right)\nonumber \\ \ldots \rho_0\left(x_n,x';\frac{\beta}{n+1}\right).
\end{eqnarray} 
of a short-time approximation of the type
\begin{eqnarray}
\label{eq:II16} \nonumber
\rho_0(x,x';\beta) = \rho_{fp}(x,x';\beta) \int_{\mathbb{R}} d \mu(a_1) \cdots \int_{\mathbb{R}} d \mu(a_q) \\ \times  \exp\left\{-\beta \sum_{i = 1}^{n_q}w_i V\left[x_r(u_i) + \sigma \sum_{k = 1}^q a_k \tilde{\Lambda}_k(u_i)\right]\right\}.
\end{eqnarray}
The quadrature points $u_i$ and weights $w_i$ as well as the functions $\tilde{\Lambda}_k(u)$ are designed such that the convergence
\[
\rho_n(x,x';\beta) \to \rho(x,x';\beta)
\]
is as fast as $O(1/n^4)$. These parameters are universal, in the sense that they are independent of the choice of potential $V(x)$, and are given in Ref.~\onlinecite{Pre04}, reference that should be consulted for further information.

At this low temperature of $100~K$, the Monte Carlo sampling requires the use of parallel tempering,\cite{Gey91, Huk96} which, however,  successfully copes with the sparse sampling problem caused by the crossing and recrossing of the barrier by the Brownian paths. As a matter of fact, by Monte Carlo integration, we compute the ratios $D_{2k}/\mathcal{N}_F$ and the associated statistical errors (two standard deviations). The quantity $\mathcal{N}_F$ is evaluated with the help of the numerical matrix multiplication technique,\cite{Kle73, Thi83a} which provides essentially exact results. Thus, the relative errors reported in Table~II are equal to the relative errors of the ratios $D_{2k}/\mathcal{N}_F$ and are, therefore, representative of the variances of the estimating functions utilized in the Monte Carlo simulation.

\begingroup
\squeezetable
\begin{table}[!htbp]
\begin{tabular}{| c | c | c | c | c | c | c | }
\hline
Order & 0 & 2 & 4 & 6 & 8 & 10 \\
\hline
Value & 5.787E-17 & 2.389E-22 &  4.010E-27 & 1.395E-31 & 7.985E-36 & 6.781E-40 \\ 
\hline
Error & $2.5\%$ & $2.4 \%$ & $2.4\%$ & $2.7\%$ & $3.9\%$ & $6.1\%$\\
\hline
\end{tabular}
\caption{\label{Tab:II} Derivatives (second row) and relative errors (third row) for the symmetric Eckart barrier at 100~K. The errors are twice the percentile relative value of the standard deviation. The errors do not include the systematic errors due to the utilization of finite-difference approximations, which have been estimated to increase the final errors with less than $2\%$. }
\end{table}
\endgroup

Once the power spectrum $\bar{G}_{F,n}(\omega)$ is determined, the absolute rate of reaction can be computed from Eq.~(\ref{eq:44}), as the quantity
\begin{eqnarray}
\label{eq:53} \nonumber
k(T)Q_r(T) && = \int_0^\infty G_{F,n}(t)dt  \\ && = \frac{1}{2} \int_{-\infty}^{\infty} G_{F,n}(t)dt = \pi \bar{G}_{F,n}(0).
\end{eqnarray}
Let us remember that
\[
G_{F,n}(t) \to G_F(t), \quad \forall t \in \mathbb{R}, 
\]
for all reconstruction algorithms that satisfy the hypothesis of Th.~\ref{Th:3}. However, as already mentioned several times, this does not automatically imply pointwise convergence in the frequency domain. Sure enough, convergence in the frequency domain is necessary only for the purpose of computing the absolute rate of reaction as the time integral of the flux autocorrelation function, the power spectrum of which is continuous at origin. It is not required for other autocorrelation functions. Because it depends on the physical significance of the corresponding autocorrelation functions and on the nature of the quantum information that is sought, the development of optimal reconstruction algorithms is a case by case problem.

It is beyond the purpose of this paper to conduct any mathematical proofs related to the pointwise convergence of the power spectrum of the flux autocorrelation functions. However, the percentile relative errors for the absolute rates of reaction presented in Table~III strongly suggest that the maximum entropy algorithm discussed in previous paragraphs is viable for the  purpose of computing rates of reaction. The errors eventually increase as the temperature is lowered, but the reader may notice that the relative errors are sufficiently small to make the algorithm useful even in the tunneling regime of temperatures ($T < 300~K$). 

At large temperatures, the relative errors converge to the relative errors for a free particle. 
The thermally-symmetrized flux autocorrelation function for the free particle is\cite{Mil83, Mil98} 
\begin{equation}
\label{eq:54}
G_F(t) = \frac{1}{\beta h} \frac{(\beta \hbar / 2 )^2}{\left[t^2 + (\beta \hbar / 2 )^2\right]^{3/2}}.
\end{equation}
Its power spectrum reads
\begin{equation}
\label{eq:55}
\bar{G}_F(\omega) =\frac{1}{\beta h} \frac{\omega \hbar \beta}{2\pi} K_1\left(\frac{\omega \hbar \beta}{2}\right),
\end{equation}
where $K_1(x)$ denotes the respective modified Bessel function of the second kind. The function $xK_1(x)$ is continuous at origin, indeed, but its even derivatives in origin are not defined. Therefore, the function $xK_1(x)$ is not readily approximated around origin by smooth functions of the type given by Eq.~(\ref{eq:45}). Thus, for example, a useful direction for future research is to modify the default model in the the maximum entropy algorithm so that to properly account for the known high-temperature limit.

\begin{table}[!htbp]
\begin{tabular}{| c | c | c | c | c | c | c | c |} 
\hline
Order of & \multicolumn{7}{c}{Temperature} \vline \\ \cline{2-8} 
derivatives & 100~K & 200~K & 300~K & 500~K & 1000~K & 2000~K & $\infty$ \\
\hline
2 & -13.8 & -2.3 & 8.4 & -2.1 & -18.3 & -25.7 & -27.6 \\
\hline
6 & -4.9 & -0.8 & 2.5 & 1.8 & -7.7 & -15.0 & -17.1 \\
\hline
10 & -2.9 & 0.3 & 0.0 & 1.3 & -5.4 & -11.9 & -13.4 \\
\hline
\end{tabular}
\caption{\label{Tab:III} Percentile relative errors for the absolute rates of reaction computed using all derivatives up to the maximum orders of $2$, $6$, and $10$, respectively. The errors are given as functions of temperature. Whenever the minimization algorithm did not converge properly while using the maximal number of derivatives, a smaller number of derivatives has been utilized. The relative errors for the high-temperature limit are those for the free particle case (which are independent of temperature).}
\end{table}

\section{Summary and discussion}

A new technique for extracting quantum dynamical information from imaginary-time data has been proposed. The technique consists in solving a symmetric Hamburger moment problem with even-order moments related to the even-order derivatives at origin of the quantum autocorrelation function. It has been demonstrated that the derivatives at origin uniquely determine the autocorrelation function. The derivatives can be computed by Monte Carlo simulations with the help of estimators of finite variance. The pointwise reconstruction of the autocorrelation functions can be performed by those inversion algorithms that satisfy the hypothesis of Th.~\ref{Th:3}, although additional care is needed if other quantities, as for instance certain integral values, are also sought. A moment based maximum entropy inversion algorithm has been numerically shown to cope successfully with the problem of computing absolute rates of reaction for a symmetric Eckart barrier. 

Perhaps, the most important step in the present development is the realization that the derivatives at origin of the imaginary-time autocorrelation functions are computable solely by Monte Carlo simulations. As argued  in the introduction, the sequence of derivatives at origin represents a set of data that is more suitable for the problem of extracting quantum dynamical information than the mere Monte Carlo evaluation of the imaginary-time autocorrelation function on a grid. However, future research is necessary in order to quantify in precise manner the efficiency of the new algorithm. In particular, the scaling of the variances of the Monte Carlo estimators with the degree of the derivatives, the dimensionality of the physical system, and the temperature must be determined. 

The numerical results presented in Section~IV demonstrate that the derivatives at origin of  autocorrelation functions contain useful information that can be utilized in at least two ways. First, one may employ this information together with various inversion algorithms for the Hamburger moment problem. In this respect, I believe that methods of Bayesian statistical inference and maximum entropy will be most useful, especially because such techniques can incorporate additional physical information (as, for instance, a certain limiting behavior) by appropriate choices of default models. Second, if only a small number of derivatives are computed,  the chemical physicist has also the option of developing certain physical models depending on parameters that can be determined from matching the known derivatives. Which of these two ways will be  the most successful for practical applications remains to be seen. 

\begin{acknowledgments} The author acknowledges support from National Science Foundation through Grant No. CHE-0096576.  He wishes to express a special thanks to Professor William H. Miller for suggestions and stimulating discussions concerning the present development. He also acknowledges an anonymous referee, whose useful comments have helped improve the presentation of the paper.
\end{acknowledgments}


\begin{thebibliography}{99}
\bibitem{Fey48} R. P. Feynman,  Rev. Mod. Phys. \textbf{20}, 367 (1948).
\bibitem{Ami93} A. M. Amini and M. F. Herman, J. Chem. Phys. \textbf{99}, 5087 (1993).
\bibitem{Ber86} B. J. Berne and D. Thirumalai, Annu. Rev. Phys. Chem. \textbf{47}, 401 (1986).
\bibitem{Cam60} R. H. Cameron,  J. Math. Phys. \textbf{39}, 126 (1960).
\bibitem{Dol84} J. D. Doll, J. Chem. Phys. \textbf{81}, 3536 (1984).
\bibitem{Fil86} V. S. Filinov, Nucl. Phys. B \textbf{271}, 717 (1986). 
\bibitem{Mak87} N. Makri and W. H. Miller, Chem. Phys. Lett. \textbf{139}, 10 (1987).
\bibitem{Dol88} J. D. Doll, T. L. Beck, and D. L. Freeman, J. Chem. Phys. \textbf{89}, 5753 (1988).
\bibitem{Mak92} C. H. Mak, Phys. Rev. Lett. \textbf{68}, 899 (1992).
\bibitem{Bay61} G. Baym and D. Mermin, J. Math. Phys. \textbf{2}, 232 (1961).
\bibitem{Nel64} E. Nelson,  J. Math. Phys. \textbf{5}, 332 (1964).
\bibitem{Sch85} H.-B. Sch\"uttler and D. J. Scalapino, Phys. Rev. Lett. \textbf{55}, 1204 (1985).
\bibitem{Whi89} S. R. White, D. J. Scalapino, R. L. Sugar, and N. E. Bickers, Phys. Rev. Lett. \textbf{63}, 1523 (1989).
\bibitem{Jar89} M. Jarrell and O. Biham, Phys. Rev. Lett. \textbf{63}, 2504 (1989).
\bibitem{Jar96} M. Jarrell and J. E. Gubernatis, Phys. Rep. \textbf{269}, 133 (1996).
\bibitem{Gub91} J. E. Gubernatis, M. Jarrell, R. N. Silver, and D. S. Sivia, Phys. Rev. B \textbf{44}, 6011 (1991).
\bibitem{Thi83} D. Thirumalai and B. J. Berne, J. Chem. Phys. \textbf{79}, 5029 (1983).
\bibitem{Gal94} E. Gallicchio and B. J. Berne, J. Chem. Phys. \textbf{101}, 9909 (1994).
\bibitem{Kim97} D. Kim, J. D. Doll, and J. E. Gubernatis, J. Chem. Phys. \textbf{106}, 1641 (1997). 
\bibitem{Kim98} D. Kim, J. D. Doll, and D. L. Freeman, J. Chem. Phys. \textbf{108}, 3871 (1998).
\bibitem{Kri01} G. Krilov, E. Sim, and B. J. Berne, J. Chem. Phys. \textbf{114}, 1075 (2001).
\bibitem{Mil83} W. H. Miller, S. D. Schwartz,  and J. W. Tromp, J. Chem. Phys. \textbf{79}, 4889 (1983).
\bibitem{Rab00} E. Rabani, G. Krilov, and B. J. Berne, J. Chem. Phys. \textbf{112}, 2605 (2000).
\bibitem{Sim01} E. Sim, G. Krilov, B. J. Berne, J. Phys. Chem. A \textbf{105}, 2824  (2001).
\bibitem{Rab02} E. Rabani, D. R. Reichman, G. Krilov, B. J. Berne, P. Natl. Acad. Sci. USA \textbf{99}, 1129 (2002).
\bibitem{Whi91} S. R. White, Phys. Rev. B \textbf{44}, 4670 (1991).
\bibitem{Whi92} S. R. White, Phys. Rev. B \textbf{46}, 5678 (1992).
\bibitem{Caf92} M. Caffarel and D. M. Ceperley, J. Chem. Phys. \textbf{97}, 8415 (1992).
\bibitem{Dei92} J. Deisz, K.-H. Luk, M. Jarrell, and D. L. Cox, Phys. Rev. B \textbf{46}, 3410 (1992).
\bibitem{Dei93} J. Deisz, M. Jarrell, and D. L. Cox, Phys. Rev. B \textbf{48}, 10227 (1993).
\bibitem{Ski89} J. Skilling, in \emph{Maximum Entropy and Bayesian methods}, edited by J. Skilling (Kluwer, Dordrecht, 1989), p. 455.
\bibitem{Sil94} R. N. Silver and H. R{\"o}der, Int. J. of Mod. Phys. C \textbf{5}, 735 (1994).
\bibitem{Dra93} D. A. Drabold and O. F. Sankey, Phys. Rev. Lett \textbf{70}, 3631 (1993).
\bibitem{Wan94} L. W. Wang, Phys. Rev. B \textbf{49}, 10154 (1994).
\bibitem{Sil96} R. N. Silver, H. R{\"o}der, A. F. Voter, and J. D. Kress, J. of Comput. Phys. \textbf{124}, 115 (1996).
\bibitem{Sil97} R. N. Silver and H. R{\"o}der, Phys. Rev. E \textbf{56} 4822 (1997).
\bibitem{Mil74} W. H. Miller, J. Chem. Phys. \textbf{61}, 1823 (1974).
\bibitem{Mil98} W. H. Miller, J. Phys. Chem. A \textbf{102}, 793 (1998).
\bibitem{Ath02} G. A. Athanassoulis and P. N. Gavriliadis, Prob. Engng. Mech. \textbf{17}, 273 (2002). 
\bibitem{Jay57} E. T. Jaynes, Phys. Rev. \textbf{106}, 620 (1957).
\bibitem{Tag94} A. Tagliani, J. Math. Phys. \textbf{35}, 5087 (1994).
\bibitem{Pre02} C. Predescu and J. D. Doll, J. Chem. Phys. \textbf{117},  7448    (2002).
\bibitem{Pre03} C. Predescu, D. Sabo, J. D. Doll, and D. L. Freeman, J. Chem. Phys. \textbf{119}, 12119 (2003).
\bibitem{Lyn68} J. N. Lyness, Math. Comput. \textbf{22}, 352 (1968). 
\bibitem{Dol99r} J. D. Doll, M. Eleftheriou, S. A. Corcelli, and David L. Freeman, \emph{Quantum Monte Carlo Methods in Physics and Chemistry,} edited by M.P. Nightingale and C.J. Umrigar, NATO ASI Series, Series C Mathematical and Physical Sciences, Vol. X, (Kluwer, Dordrecht, 1999).
\bibitem{Ber70} B. J. Berne and G. D. Harp, Adv. in Chem. Phys. \textbf{17}, 63 (1970).
\bibitem{Ham20} H. Hamburger, Math. Ann. \textbf{81}, 235 (1920); \textbf{82}, 120 (1921); \textbf{82}, 168 (1921).
\bibitem{Dur96} R. Durrett, \emph{Probability: Theory and Examples,} 2nd ed. (Duxbury, New York, 1996).
\bibitem{Sim79} B. Simon, \emph{Functional Integration and Quantum Physics} (Academic, London, 1979).
\bibitem{Bar79} J. Barker, J. Chem. Phys. \textbf{70}, 2914 (1979). 
\bibitem{Her82} M. F. Herman, E. J. Bruskin, and B. J. Berne, J. Chem. Phys. \textbf{76}, 5150 (1982).
\bibitem{Pal33} R. Paley, N. Wiener, and A. Zygmund, Math. Z. \textbf{37}, 647 (1933).
\bibitem{Pre03c} C. Predescu, D. Sabo, J. D. Doll, and D. L. Freeman, J. Chem. Phys. \textbf{119}, 10475 (2003).
\bibitem{Kle73} A. D. Klemm and R. G. Storer, Aust. J. Phys. \textbf{26}, 43 (1973).
\bibitem{Thi83a} D. Thirumalai, E. J. Bruskin, and B. J. Berne, J. Chem. Phys. \textbf{79}, 5063 (1983). 
\bibitem{Yam04} T. Yamamoto and W. H. Miller, J. Chem. Phys. \textbf{120}, 3086 (2004).
\bibitem{Jay78} E. T. Jaynes, in \emph{The Maximum Entropy formalism,} edited by R. D. Levine and M. Tribus (MIT Press, Cambridge, 1978), pp.~15-118.
\bibitem{Ski89a} J. Skilling, in \emph{Maximum Entropy and Bayesian methods}, edited by J. Skilling (Kluwer, Dordrecht, 1989) p. 45.
\bibitem{Gul89a} S. F. Gull, in \emph{Maximum Entropy and Bayesian methods}, edited by J. Skilling (Kluwer, Dordrecht, 1989) p. 53.
\bibitem{Tag98} A. Tagliani, J. Comput. Appl. Math. \textbf{90}, 157 (1998).
\bibitem{Mil03} W. H. Miller, Y. Zhao, M. Ceotto, and S. Yang, J. Chem. Phys. \textbf{119}, 1329 (2003).
\bibitem{Hid01} Y. Hida, X. S. Li, and D. H. Bailey, in Proceedings of the 15th IEEE Symposium on Computer Arithmetic, IEEE Computer Society, 2001, pp. 155-162.
\bibitem{Han94} N. F. Hansen and H. C. Andersen, J. Chem. Phys. \textbf{101}, 6032 (1994).
\bibitem{Kha99} I. R. Khan and R. Ohba, J. Comput. Appl. Math. \textbf{107}, 179 (1999).
\bibitem{Pre04} C. Predescu, Phys. Rev. E \textbf{69}, 056701 (2004).
\bibitem{Gey91} C. J. Geyer, in \emph{Computing Science and Statistics: Proceedings of the 23rd Symposium on the Interface,} edited by E. M. Keramigas, (Interface Foundation: Fairfax, 1991), pp. 156 - 163.
\bibitem{Huk96} K. Hukushima and K. Nemoto, J. Phys. Soc. Jpn. \textbf{65}, 1604 (1996).
\end{thebibliography}
\end{document}